\documentclass[useAMS,usenatbib]{mn2e}

\usepackage{graphicx}

\title[The nature of the infrared counterpart of IGR J19140+0951]{The nature of the infrared counterpart of\\
 IGR J19140+0951}
\author[D.C. Hannikainen et al.]{D.C. Hannikainen$^{1}$\thanks{E-mail:
diana@astro.helsinki.fi}, M.G. Rawlings$^{2}$, P. Muhli$^{1}$, O. Vilhu$^{1}$, J. Schultz$^{1}$
and 
\newauthor
J. Rodriguez$^{3}$ \\
$^{1}$Observatory, PO Box 14, FI-00014 University of Helsinki, Finland\\
$^{2}$Joint Astronomy Centre, 660 N. A'ohoku Place, University Park, Hilo, Hawaii, 96270 U.S.A.\\
$^{3}$AIM -- Astrophysique Interactions Multi-\'echelles (UMR 7158 CEA/CNRS/Universit\'e Paris 7 Denis Diderot), CEA Saclay, \\
DSM/DAPNIA/Service d'Astrophysique, B\^at. 709, l'Orme des Mesuriers, FR-91 191 Gif-sur-Yvette Cedex, France
}
\begin{document}

\date{Accepted . Received ; in original form}

\pagerange{\pageref{firstpage}--\pageref{lastpage}} \pubyear{2002}

\maketitle

\label{firstpage}

\begin{abstract}
The INTEGRAL observatory has been (re-)discovering new X-ray sources
   since the beginning of nominal operations in early 2003. 
   These sources include X-ray binaries, Active Galactic Nuclei, cataclysmic variables, etc.
Amongst the X-ray binaries, the true nature of many of these sources has remained largely elusive, though they seem to 
   make up a population of highly absorbed high-mass X-ray binaries. 
   One of these new sources,  IGR~J19140$+$0951, was serendipitously
   discovered on 2003 Mar 6 during an observation of the galactic microquasar GRS 1915$+$105.
   We observed IGR~J19140$+$0951 with UKIRT in order to identify the infrared counterpart.
  Here we present the H- and K-band spectra.
   We determined that the companion is a B0.5-type bright supergiant  in a wind-fed
   system, at a distance $\la$ 5 kpc.
\end{abstract}


\section{Introduction}
The INTErnational Gamma-Ray Astrophysics Laboratory ({\it INTEGRAL}, Winkler et al. 2003), launched on 2002 October 17 and beginning normal operations in early 2003, has
been discovering (or in some cases re-discovering) many new X-ray and gamma-ray sources.
The first such source, discovered with the IBIS/ISGRI instrument (Lebrun et al. 2003; Ubertini et al. 2003) during a galactic plane scan, was designated IGR~J16318$-$4848 (Walter et al. 2003). 
The infrared/optical counterpart indicates that IGR~J16318$-$4848 is probably a high-mass X-ray binary.
Furthermore, the source is observed to be intrinsically strongly absorbed by cold matter. 
Subsequently, more than two hundred new sources have been discovered to date (e.g. Bodaghee et al. 2007; Bird et al. 2007), the majority being low-mass or high-mass X-ray binaries, and Active Galactic Nuclei.
Amongst the high-mass X-ray binaries, a large fraction of the sources are found to be heavily obscured sources, exhibiting much larger column densities than would be expected along the line of sight (e.g. Kuulkers 2005). These systems are likely to be supergiant high-mass X-ray binaries, largely missed in previous X-ray surveys which scanned the skies with instruments geared more towards the softer X-ray energies.

One of the new sources discovered with {\it INTEGRAL} is IGR~J19140$+$0951 \citep{hannikainen03,  hannikainen04}.
Inspection of the high energy archives showed it to be the most likely 
   hard X-ray counterpart to the poorly studied {\it EXOSAT} source
   EXO~1912$+$097 \citep{lu96}. 
A target of opportunity was performed on IGR~J19140$+$0951 with the {\it Rossi
  X-Ray Timing Explorer (RXTE)} -- preliminary analysis showed the source 
   had a rather
   hard spectrum, fitted with a power law of photon index 1.6 and an 
   absorption column density $N_H = 6\times10^{22} \rm cm^{-2}$
   \citep{swank03}.
Timing analysis of the {\it RXTE}/ASM data revealed an X-ray period 
   of 13.55 days \citep{corbet04}.
This implies that the source was detected even in the early days of 
   the {\it RXTE} mission, which in turn suggests that IGR~J19140$+$0951 is a 
   persistent source although most of the time in the faint state. 

   \begin{figure*}
   \centering
   \includegraphics[width=9cm,angle=270]{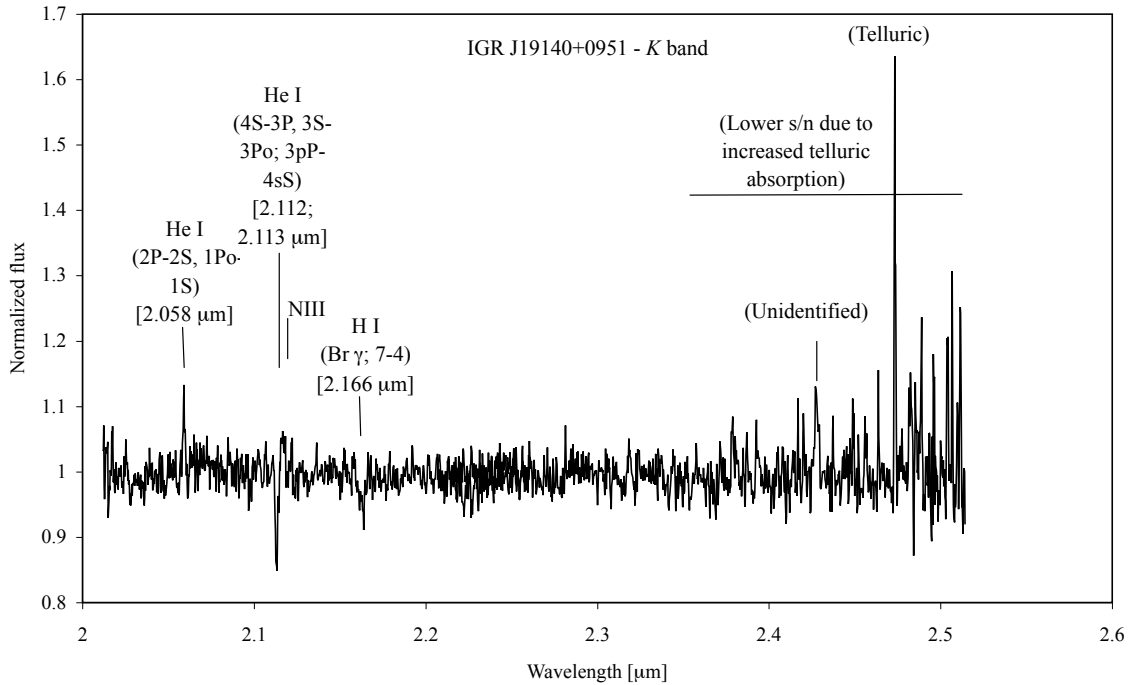}
   \caption{Normalized short- and long-K spectra of the infrared
counterpart of IGR~J19140$+$0951. Spectroscopic features
constraining the spectral type are indicated. The CO
bandheads characteristic of late-type stars are clearly
absent (e.g. at 2.2935, 2.3227, 2.3535 $\mu$m). 
Strong emission features longward of 2.35$\mu$m
are due to imperfect cancellation of telluric bands with
the standard.}
              \label{fig1}%
    \end{figure*}

   \begin{figure*}
   \centering
   \includegraphics[width=10cm,angle=270]{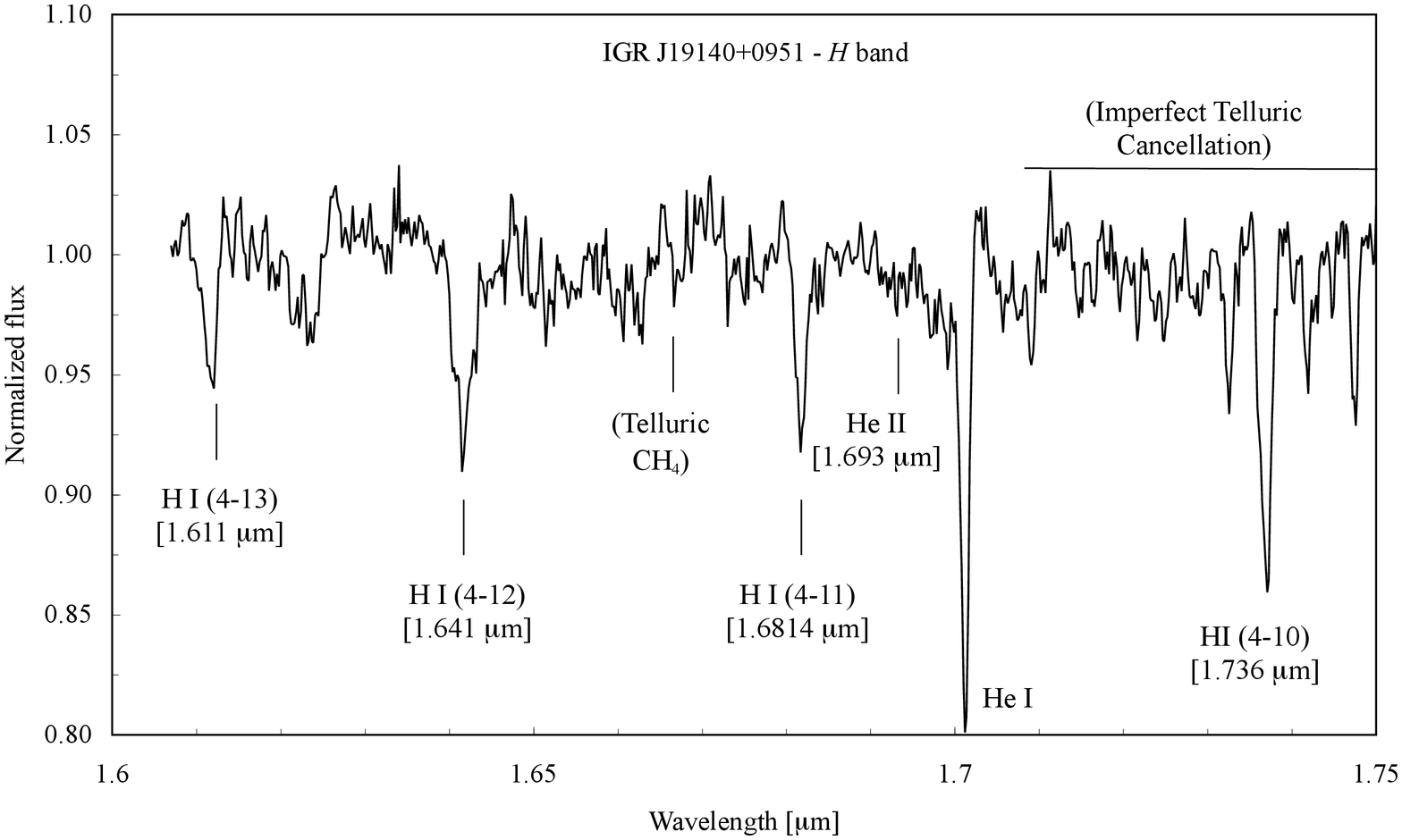}
   \caption{Normalized H-band spectra of the infrared
counterpart of IGR~J19140$+$0951.}
              \label{fig2}%
    \end{figure*}

High energy spectral analysis of IGR~J19140$+$0951 covering the period of its
   discovery (i.e. {\it INTEGRAL} revolution 48) was presented in 
   Hannikainen et al. (2004).
During this observation, the source, although very variable, showed two
   distinct spectral behaviours. 
The first one manifests a thermal component (blackbody-like) in the soft
   X-ray and a hard X-ray tail, while the second one is harder and can 
   be interpreted as originating from thermal Comptonization. 

In a follow-up paper, we reported on high energy observations performed
   with {\it RXTE} on three separate occasions: in 2002, 2003 and again
   in 2004 \citep{rodriguez05}.
We also analyzed {\it INTEGRAL} observations from the period early March
   to mid-May 2003. 
We showed that IGR~J19140$+$0951 spends most of its time in a very low-luminosity
   state characterized by thermal Comptonization.
On some occasions, we observed the luminosity to vary by a factor of 10 during 
   which the spectrum can show evidence of a thermal component, besides
   thermal Comptonization by a hotter plasma than during the low luminosity
   state. 
The spectral parameters derived strongly pointed to IGR~J19140$+$0951 hosting 
   a neutron star, although a black hole could not be ruled out, and 
   we concluded that IGR~J19140$+$0951 is most probably a High Mass X-ray Binary
   (HXMB).

{\it Swift} observations of the source caught it at its lowest luminosity level thus far: 
   about five times fainter than the dimmest observation published in Rodriguez et al. (2005)
    and about 150 times fainter than the brightest state of the source seen with {\it INTEGRAL}
    (Rodriguez et al. 2006).

The infrared counterpart to IGR~J19140$+$0951 was identified by
   in't Zand et al. (2006), following an accurate position determination using
   {\it Chandra}, with the heavily reddened 2MASS 19140422$+$0952577
   in the 2MASS catalogue.
Recently, Nespoli, Fabregat \& Mennickent (2007), based on K-band observations,
   estimate the spectral type of the infrared counterpart to be B1 I.

In this paper, we present both H- and K-band infrared data obtained 
    with the UK Infrared Telescope (UKIRT) in Hawaii of IGR~J19140$+$0951.
In Section 2 we describe the observations and the data reduction, while 
   in Section 3 we discuss the outcome of the observations.

\section[]{Infrared observations}

\subsection{K-band}
Near-infrared observations were carried out on the night of
2006 March 7 (UT) using UIST on the 3.8-m UKIRT at the 
Mauna Kea Observatory in Hawaii. 
The full 1024$\times$1024-pixel array was used in conjunction
with the short- and long-K grisms. A 4-pixel slit at a
position angle of 12.3 degrees East of North was used in
order to enable simultaneous spectroscopy of both the
IGR J19140+0951 infrared counterpart and the nearby brighter
star for comparison. A standard 12'' nod along the slit
was used. This configuration produced a total overall
effective wavelength range of 2.0--2.5 $\mu$m.
The weather was clear, with seeing averaging around 0.4''.
Suitable flat field frames were taken, and the data were
flat fielded by the data reduction pipeline during the
observations. Spectra of the F6V standard BS 7354 were
taken at matching airmasses to facilitate flux calibration
and cancellation of telluric absorption.

Data were reduced using the Starlink packages Figaro, Kappa
and Gaia. One-dimensional spectra were extracted from the
coadded sky-subtracted frames and wavelength calibrated
against argon lamp spectra. Br$\gamma$ absorption
was interpolated out of the standard star spectrum. This
was done in order to avoid introducing contamination into
the target spectra when dividing by the standard to remove telluric effects.
Where necessary, cosmic ray spikes in the resultant 
spectra were interpolated out, and the data were linearly
rebinned and flux calibrated against the known properties
of the standard. The fully reduced spectra were then divided
by fitted polynomial continua in order to produce the
normalised spectrum shown in Fig. 1. 
The polynomials are established interactively using the
Starlink Figaro CFIT routine, and typically $\sim$20 points
are used to constrain the fit for spectra of this type.
The selection of the continuum positions used is performed
with reference to published stellar and telluric line lists
to minimize contamination. If not simply an
artifact of imperfect telluric cancellation, the feature at
2.42--2.43 $\mu$m remains unidentified. It unfortunately
falls between the wavelength ranges covered by  
infrared atlases, including Hanson et al. (2005). 
However, the possible presence of this feature does not
affect the classification of the star. 

\subsection{H-band}

H-band spectroscopy data were acquired on 2006 October 5 (UT) using the UKIRT/UIST long-H grism with a 4-pixel slit in second order.  The same pointing, nod and slit orientation were used as for the K-band data.  The total on-source integration time was 240s.  One-dimensional spectra were extracted from the flat-fielded, coadded frames using the same method and data reduction software as for the K-band data, and wavelength-calibrated against an argon lamp spectrum.  Although a standard star spectrum was obtained just after that of the science target, this was later found to be of insufficient quality, and so long-H observations of HD 122563 (F8IV) taken on 2005 January 30 (UT) were used instead. Extraneous photospheric lines were identified and interpolated out of the normalised standard spectrum.

Before dividing the normalised spectrum of IGR J19140$+$0951 with that of the replacement standard, a correction for airmass difference was also necessary. Telluric absorption features in the standard were therefore logarithmically rescaled to produce optimal cancellation in the spectrum of IGR J19140$+$0951 (see Sec. 3.1 of Rawlings et al. 2003 for a full description of this technique).  To test the possible impact of the telluric cancellation method on key lines, the logarithmic scaling factor was varied to produce over- and under-cancellation.  It was found that the equivalent width ratios of the photospheric lines used for classification by Hanson et al. (2005) varied by no more than 2 per cent. The effect of the airmass correction on the classification of the IR counterpart of IGR J19140$+$0951 was therefore deemed negligible.

\section{Results}

In this section we summarize the results based on the K- and H-band spectra.

\subsection{K-band spectrum}

Figure~1 shows the UKIRT short- and long-K spectrum of IGR~J19140$+$0951
  (or strictly speaking, of 2MASS 19140422$+$0952577, the star associated
  with the X-ray position of IGR~J19140$+$0951). 
In order to identify the stellar type of the companion to IGR~J19140$+$0951, we used the 
  infrared atlases of Hanson et al. (2005) and Hanson, Conti \& Rieke (1996). 
The spectral lines we used in the identification were the He\,{\sc i} lines at 2.0581 
   and 2.1126 $\mu$m,
  the N\,{\sc iii} line at 2.1155  $\mu$m, and the H\,{\sc i} (Br-$\gamma$) line at 2.1661 $\mu$m
  which are all marked in Fig.~1. 
The first thing to note is that the first He\,{\sc i} line and N\,{\sc III} are in emission, while the second 
   He\,{\sc i} line  and H\,{\sc i} are in absorption,
  which is what is expected for a hot and luminous early-B supergiant.
These are the same spectral features that were used by Nespoli et al. (2007)
   in their classification.
The fact that the He\,{\sc i} line at 2.0581 $\mu$m is in emission is probably a luminosity effect
   (Hanson et al. 1996).
The relative strengths of the He\,{\sc i}  and H\,{\sc i} point to a B0.5 or B1 star.
Also, the H\,{\sc i} line appears to be a blend of He\,{\sc i} at 2.161  $\mu$m and
  the H\,{\sc i} itself.  
In addition, the N\,{\sc iii} line in emission is most prominent in the supergiant classes, 
  i.e. luminosity class Ia or Iab. 
Hanson et al. (2005) mention that the N\,{\sc iii} line could also be C\,{\sc iii}.

We would like to point out that, similarly to Hanson et al. (1996), we did not use
   equivalent widths in our classification, as there are substantial variations between 
   stars -- we have not conducted an  absolute classification but a comparison between spectra. 

\subsection{H-band spectrum}
   
Figure~2 shows the H-band spectrum of IGR~J19140$+$0951.
The lines used in identifying the source are marked in the figure, the most prominent line being the He\,{\sc i} line at 1.7002 $\mu$m in absorption. 
Based on the narrowness of the lines, especially of the He\,{\sc i} line, we can confirm that the source is indeed a supergiant. 
A study of the profile of the He\,{\sc i} line near 1.7 $\mu$m suggests the wings
are possibly real, rather than merely a cancellation artifact. We speculate 
that this may indeed be due to the stellar wind.
By comparing the relative strengths of the He\,{\sc i}  and H\,{\sc i} lines in our spectrum with those of Hanson et al. (2005), we narrow down our classification and say that we are dealing with a B0.5 star.
So we conclude that the companion to IGR~J19140$+$0951 is a bright supergiant
   in the B0.5  class. 

\subsection{IGR J19140+0951 as an HMXB}

A B0.5  supergiant counterpart identifies  IGR~J19140$+$0951 as an HMXB.
Most HMXB's belong to the class of Be-binaries (Liu et al. 2000) --
 as we do not see the H\,{\sc i} (Br-$\gamma$) line in emission, we
  can say that IGR~J19140$+$0951 does not belong to this class, but instead belongs
  to the relatively underpopulated class of normal early B-type binaries. 
Indeed, of all the known HMXB's, only 25\% belong to the class of supergiants
  (Charles \& Coe 2006).
The only two known HMXB's with a B0.5 star are SMC X-1 and Vela X-1, and
   both of these are subluminous supergiants (Liu et al. 2000).
They also both contain pulsars.
Other pulsars are in binary systems with either normal giants or even main sequence
  stars.
  No X-ray pulsations have been detected from IGR~J19140$+$0951. 
  
A B0 supergiant has a mass of 25 M$_{\sun}$ and a radius of 30 R$_{\sun}$
    (Cox 2000, Table 15.8).
  Assuming a period of 13.55 days and a neutron star of mass 1.4 M$_{\sun}$,
  the orbital separation is a=71.15 R$_{\sun}$ and the B supergiant's Roche lobe
  is about 44 R$_{\sun}$. 
  Hence, the star is inside the Roche lobe and the system is wind-fed as presumed.
  If one replaces the neutron star with a 10 solar mass black hole, the Roche lobe
  radius of the B supergiant is then 35 R$_{\sun}$ and then the system would still remain
  purely wind-fed.

So far, of  the $\sim$200 new sources discovered  with  {\it INTEGRAL} (those designated   with ``IGR'') 50\% are as yet unclassified.
 Six sources, or 3\%, are LMXB's, while another 32 sources are HMXB's (Bodaghee 2007).
  Of these HMXB's, the majority belong to the class of heavily obscured sources (e.g. Kuulkers 2005) with OB supergiant companions. 
  The absorption column to IGR~J19140$+$0951 varied between $5\times10^{22}$ -- $1\times10^{23}$ cm$^{-2}$ (Rodriguez et al. 2005)
making it one of the heavily obscured sources, if one takes  $2\times10^{22}$ cm$^{-2}$ 
   to be the galactic column density.
  Most of the obscured sources can be found in the Norma arm of the Milky Way,
whereas IGR~J19140$+$0951 lies in the direction of the tangent to the Sagittarius arm. 
The abundance of heavily obscured sources in the Norma arm may be the result of it being the site with the highest formation of rate of OB stars (e.g. Bronfman et al. 1996). 
When coupled with a compact object to form an X-ray binary, these supergiant systems are wind-fed accretors, and therefore are likely to be heavily obscured.
In contrast, the Sagittarius arm has a lower OB star formation rate, and hence will contain fewer of these absorbed sources. 
In fact, the Norma arm is closer to the Galactic Centre than the Sagittarius arm, and its proximity to the 3-kpc molecular ring makes it a region that is naturally more dense with material favouring an enhanced birth rate of OB stars. 
   As mentioned in in't Zand et al. (2006), if the association of IGR~J19140$+$0951 with the Sagittarius arm is true, then the distance
 is of the order 2--6 kpc, and the implied 1--20 keV {\it Chandra} luminosity is then
 $10^{35}$ erg s$^{-1}$ which is common for HMXB's. 
  In fact, for a typical B0 supergiant the absolute K-magnitude is $M_K$=$-5.8$ (Cox 2000, Tables 7.8 and 15.7) and for IGR~J19140$+$0951 the K extinction $A_K$ is about 1 magnitude (Cox 2000, Table 21.6) assuming $A_V$=11 as found in in't Zand et al.  (2006). 
  They also measured the apparent K-magnitude to be $m_K$=8.7. 
  From these numbers it follows (M=m+5$-$5 log d $-$ A, distance d in pc) that the distance is 5 kpc.
  Using B0.5 I instead of B0 I would give a slightly smaller distance.
  This supports placing the source in the Sagittarius arm.

\section{Conclusions}

We have identified the counterpart to IGR~J19140$+$0951 as being a B0-B1 supergiant,
 with the most favoured type being a bright B0.5 supergiant, consistent with the estimate of Nespoli et al. (2007).
This implies that  IGR~J19140$+$0951 belongs to the class of HMXB's.
In addition, the absorbing column to IGR~J19140$+$0951 suggests that it also belongs
  to the class of newly discovered heavily obscured {\it INTEGRAL} sources. 
We estimated a distance of $\la$ 5 kpc, placing it in the Sagittarius arm.

\section*{Acknowledgments}
      DCH gratefully acknowledges a Fellowship from the Academy of Finland.
      The authors wish to thank Arash Bodaghee and
      JR wishes to thank C. Gouiffes for fruitful discussions.
      MGR wishes to thank Andy Adamson for helpful discussions and advice. 
      We thank the anonymous referee for useful comments.
      We are very grateful to the staff at UKIRT for their assistance in obtaining
       the data presented in this paper. The data presented here were obtained 
       during Director's Discretionary Time. UKIRT, the United Kingdom Infrared Telescope,
       is operated by the Joint Astronomy Centre on behalf of the UK Particle Physics
       and Astronomy Research Council.

\bsp

\label{lastpage}

\end{document}